\documentstyle[aps]{revtex}
\newcommand{\D}{\mbox{d}}

\renewcommand{\Re}{\mbox{Re}}
\newcommand{\ve}{\varepsilon}
\begin{document}
\draft
\title{Tunneling through ultrasmall NIS junctions in terms of Andreev
reflection: \protect\\ a nonlinear response approach}

\author{A. H{\"a}dicke and W. Krech}
\address{Institut f{\"u}r Festk{\"o}rperphysik,
        Friedrich-Schiller-Universit{\"a}t Jena,
        Max-Wien-Platz 1\\
        D--07743 Jena, Germany}
\date{April 1995}
\maketitle
\begin{abstract}
The Andreev current through an ultrasmall NIS junction is calculated
in a systematic way by means of a nonlinear response approach basing
on the elementary Hamiltonian of quasiparticle tunneling. The voltage
dependence of current and differential conductance as well as the
Andreev conductance are derived for low- and high-impedance environments,
respectively.
\end{abstract}

\section{Introduction}
\label{sec1}

Effects of single-charge tunneling at junctions with
ultrasmall capacitances have been studied both theoretically and
experimentally during the last years. For a review see \cite{gra2}.
Modern nanolithography allows the fabrication of tunnel junctions with
capacitances less than $C<10^{-16}$F which means that thermal fluctuations
can be disregarded at the 1K scale. Since the tunnel resistances $R$
satisfy the condition $R\gg R_Q$ quantum fluctuations can be neglected.
$R_Q$ is the quantum resistance $R_Q=h/e^2$. Different types of
junctions and several charge transport mechanisms depending on whether
the electrodes are normal or superconducting ones have been investigated.

Here the simple example of a voltage driven NIS junction in series with an
environment resistance $R_E$ is studied at zero
temperature. At $T=0$ there is no charge transport through a NIS junction
in terms of quasiparticles for voltages lower than the threshold
${\mit\Delta}/e$, where $2{\mit\Delta}$ is the energy gap of the
superconductor (${\mit\Delta}\equiv {\mit\Delta}(0)$). The reason is that
there are
no quasiparticles in the superconductor with energies below ${\mit\Delta}$.
The quasiparticle current $\langle I\rangle_{qp}$ reads \cite{tin2}
\begin{equation}
\langle I\rangle_{qp}(V)=
\frac{1}{R}\sqrt{V^2-\left(\frac{{\mit\Delta}}{e}\right)^2}
{\mit\Theta}\left(V-\frac{{\mit\Delta}}{e}\right)-[V\to -V]\;,
\label{qpc1}
\end{equation}
where the symbol ${\mit\Theta}$ stands for the unit step function.
Eq.~(\ref{qpc1})
is valid at $T=0$ and in case of a vanishing environment resistance
($R_E/R_Q\to 0$). Since the electromagnetic environment is able to absorb
energy it influences the current-voltage characteristic of a single junction
essentially. In the limit of a high-resistance environment
($R_E/R_Q\to\infty$) the voltage threshold ${\mit\Delta}/e$ is increased by
the Coulomb blockade $E_c/e=e/(2C)$. The corresponding expression of the
current can be get from Eq.\ \ref{qpc1} by the replacement
$V\to V-E_c/e$;
\begin{equation}
\langle I\rangle_{qp}(V)=
\frac{1}{R}\sqrt{\left(V-\frac{E_c}{e}\right)^2-
\left(\frac{{\mit\Delta}}{e}\right)^2}
{\mit\Theta}\left(V-\frac{E_c}{e}-\frac{{\mit\Delta}}{e}\right)-[V\to -V]\;.
\label{qpc2}
\end{equation}
But at the moment let us ignore the environment. There is another charge
transport mechanism leading to a nonvanishing current in the the subgap
region ($0<V<{\mit\Delta}/e$). It corresponds to the transfer of two
electrons which can be converted into a Cooperpair instantaneously. This
process, known as Andreev reflection in case of no barrier for a long time
\cite{and3}, exists
under subgap conditions because it avoids the generation of excited states
in the superconductor. In this connection the influence of imperfections
in the barrier or the electrodes as well as the proximity effect is
disregarded. The current which is much smaller than ordinary quasiparticle
tunneling (Eq.~\ref{qpc1}) corresponds to a process of higher order in
perturbation theory and the conductivity emerges to be proportional to
$R^{-2}$. The effect was described by means of the Bogoliubov-de Gennes
equations \cite{blo1} as well as in terms of a transition rate formulation
\cite{hek3,hek2,hek1,hes1,han1}. Here, we are going to show how the nonlinear
response approach basing on the elementary Hamiltonian of quasiparticle
tunneling generates in a systematic and direct way those current contributions
(beside others) which correspond to the Andreev reflection. After sketching
the model (Sec.\ \ref{sec2}) and basic concepts of nonlinear response
theory (Sec.\ \ref{sec3}), we derive the Andreev current (Secs.\
\ref{sec4} and \ref{sec5}) for the low-impedance and the high-impedance cases
(Sec.\ \ref{sec7}), respectively. We restrict ourselves to the concept of
ballistic motion of electrons. Finally, a simple approximation in case of
the high-impedance environment is given (Sec.\ \ref{sec7}).

\section{The model}
\label{sec2}
The Hamiltonian of a voltage driven NIS junction with
environment reads
\begin{equation}
\label{ham1}
H = QV+H_{res}+H_T\;,
\end{equation}
where the tunnel Hamiltonian is given by
\begin{eqnarray}
H_T&=&H_++H_-\;,\qquad \qquad H_-=H_+^{\dagger}\;,\nonumber\\
\label{ham2}
H_+&=&\sum\limits_{l,r,\sigma}T_{lr}c^{\dagger}_{r,\sigma}c_{l,\sigma}
e^{i{\mit\Phi}}\;.
\end{eqnarray}
$c_{l,\sigma}$ and $c_{r,\sigma}$ stand for electron annihilation
operators of the
left and right electrode, respectively, satisfying anticommutation relations.
The spin is labeled by the subscript $\sigma$. The sum is taken over
momenta $l$ and $r$ and the spin ($\sigma=\uparrow,\downarrow$).
The reservoir Hamiltonian $H_{res}=H_L+H_R+H_E$ consists of terms
corresponding to the left ($L$) and right ($R$) electrodes and the
environment ($E$) which can be described in standard way \cite{dev1,gra1}.
Owing to the phase operator tunneling is connected with excitations
in the electromagnetic environment. The convention is chosen in such a way
that a positive voltage favors tunneling from left to right which reduces
the junction charge $Q$ by $e$.
The operator $H_+$ means tunneling from left to right in contrast
to the Hermitian conjugate which describes the reverse process.
$T_{lr}$ are the tunneling matrix elements.
The basic algebra ruling this approach is the relation
\cite{ave1}
\begin{equation}
\label{alg1}
H_{\pm}\cdot F(Q)=F(Q\pm e)\cdot H_{\pm}\:,
\end{equation}
where $F$ is an arbitrary function of the junction charge
$Q$. The operator $\exp(\pm i{\mit\Phi})$ changes the macroscopic charge
$Q$ on the junction by the value $\mp e$.
The algebra (\ref{alg1}) corresponds to the elementary commutation relation
\[[Q,{\mit\Phi}]=ie\;.\]

An important point is to take into account the specific feature of a NIS
junction. Without loss of generality we assume that the right electrode
is the superconducting one. Therefore, Bogoliubov
transformed operators are introduced on the right-hand side in the following
way
\begin{eqnarray}
c_{r,\sigma}&=&u_{r,\sigma}\gamma_{r,\sigma}+
v_{r,\sigma}\gamma^{\dagger}_{-r,-\sigma}\;,\nonumber\\
c^{\dagger}_{r,\sigma}&=&u_{r,\sigma}\gamma^{\dagger}_{r,\sigma}+
v_{r,\sigma}\gamma_{-r,-\sigma}\;,
\label{bog}
\end{eqnarray}
where $\gamma_{r,\sigma}$ and $\gamma^{\dagger}_{r,\sigma}$ are new
Fermi operators satisfying common
anticommutation relations and the numerical coefficients $u_{r,\sigma}$ and
$v_{r,\sigma}$ are the known BCS coherence factors \cite{tin2}. We remind of
the off-diagonal Hamiltonian of the superconducting electrode which becomes
diagonal in the new operators $\gamma^{\dagger}$ and $\gamma$
\begin{equation}
H_R=\mbox{const}+
\sum\limits_{r,\sigma}E_r\gamma^{\dagger}_{r,\sigma}\gamma_{r,\sigma}\;.
\label{ham3}
\end{equation}
The quasiparticle energies of the superconductor read
\begin{equation}
E_r=\sqrt{\ve_r^2+{\mit\Delta}^2}\;.
\label{eng1}
\end{equation}
$H_L$ on the left-hand side is given by
\begin{equation}
H_L=\sum\limits_{l,\sigma}\ve_l c^{\dagger}_{l,\sigma}c_{l,\sigma}\;,
\label{ham4}
\end{equation}
where $\ve_l$ are the usual electron energies counted with respect to the
Fermi energy.

\section{Nonlinear response theory}
\label{sec3}
The dynamics of our physical system modeled by
an unperturbed Hamiltonian $H_o$ and an interaction (tunneling) $H_T$ is
described by the statistical operator $\rho$ satisfying the von Neumann
equation which corresponds in the interaction representation (superscript
$(I)$) to the integral equation
\begin{equation}
\label{neq1}
\rho(t)^{(I)}=\rho_o-\frac{i}{\hbar}\int\limits_{-\infty}^{t}
[H_T^{(I)}(t'),\rho^{(I)}(t')]\D t'\;.
\end{equation}
It is assumed that the interaction is switched on at $t=-\infty$
adiabatically. $\rho_o$ is given by the canonical expression
($\beta=1/(k_BT)$)
\[
\rho_o=\frac{e^{-\beta H_o}}{\mbox{tr}\{e^{-\beta H_o}\}}\;.
\]
The mean value of an arbitrary operator can be calculated in terms of a
successive approximation. In case of the current operator
defined by
\begin{equation}
I=-\frac{\D}{\D t}Q=-\frac{1}{i\hbar}[Q,H]
=\frac{e}{i\hbar}(H_+-H_-)
\label{cop1}
\end{equation}
one finds the expansion
\begin{eqnarray}
\langle I\rangle&=&
\frac{1}{i\hbar}\int\limits_{-\infty}^{t}\D t'\,
\langle[I^{(I)}(t),H_T^{(I)}(t')]\rangle_o\nonumber\\
\label{ac1}
&&+\left(\frac{1}{i\hbar}\right)^3\int\limits_{-\infty}^{t}\D t'
\int\limits_{-\infty}^{t'}\D t'\int\limits_{-\infty}^{t''}\D t'''\,
\langle[[[I^{(I)}(t),H_T^{(I)}(t')],H_T^{(I)}(t'')],H_T^{(I)}(t''')]\rangle_o
+\ldots
\end{eqnarray}
A term of zeroth order is missing because in case of no
interaction (tunneling) there is also no current.
Now the quasiparticle current according to the first order term in
Eq.~(\ref{ac1}) can be calculated. The result
\begin{equation}
\label{qpc3}
\langle I\rangle_{qp}=-\frac{2e}{\hbar^2}\Re\int\limits_{-\infty}^t\D t'\,
\langle[H_+^{(I)}(t),H_-^{(I)}(t')]\rangle_o
\end{equation}
leads at $T=0$ to the Eqs.~(\ref{qpc1}) and (\ref{qpc2}), respectively.
The Andreev current is contained in the following terms of second
nonvanishing order
\begin{eqnarray}
\langle I\rangle=\frac{2e}{\hbar^4}\Re\int\limits_{-\infty}^t\D t'
\int\limits_{-\infty}^{t'}\D t''\int\limits_{-\infty}^{t''}\D t'''&\Big\{&
\langle[[[H_+^{(I)}(t),H_+^{(I)}(t')],H_-^{(I)}(t'')],H_-^{(I)}(t''')]
\rangle_o\nonumber\\
&&+[[[H_+^{(I)}(t),H_-^{(I)}(t')],H_+^{(I)}(t'')],H_-^{(I)}(t''')]
\rangle_o\nonumber\\
\nonumber\\
\label{ac2}
&&+[[[H_+^{(I)}(t),H_-^{(I)}(t')],H_-^{(I)}(t'')],H_+^{(I)}(t''')]
\rangle_o\Big\}\;,
\end{eqnarray}
where the angle brackets $\langle\ldots\rangle_o$ denote averaging with
respect to $\rho_o$.
In Eq.~(\ref{ac2}) only correlations with vanishing
signature ($++--$ and their permutations) are taken into account. The
reason is that the separation of the voltage dependence in the correlation
functions ($\langle\ldots\rangle_o$) with nonvanishing signature
(e.g. $\langle H_+H_+H_+H_+\rangle_o$) leads to expressions
containing time dependent ($t$) terms. Furthermore, a phase ${\mit\Phi}_o$
remains indeterminated additionally. This is like in common Josephson physics
where contributions proportional to $\sin(2eVt/\hbar+{\mit\Phi}_o)$ and
$\cos(2eVt/\hbar+{\mit\Phi}_o)$\cite{rog1} arise. However, with respect to
the Josephson effect the phase ${\mit\Phi}_o$ is adjusted by current biasing.
Therefore, in our opposite case of voltage biasing, where the charge
becomes adjusted, these terms do not contribute.

According to the fact that Cooper pairs live in the condensate the
corresponding energy balance of the Andreev current does not contain
any energy contributions belonging to the superconducting bank.
It turns out that the Andreev current is only determined by the first
summand of Eq.~(\ref{ac2}) because the energy balances of the two other
summands always depend on the quasiparticle energies $E_r$.

\section{Calculation of the correlation function}
\label{sec4}
The voltage dependence can be separated by means of Eq.~(\ref{alg1}) and
in terms of the new time variables
\[
\tau\equiv t-t';\qquad\tau'\equiv t'-t'';\qquad\tau''\equiv t''-t'''
\]
one gets the Andreev current $\langle I\rangle^{(A)}$
\begin{equation}
\langle I\rangle^{(A)}=\frac{2e}{\hbar^4}\Re\int\limits_0^{\infty}\D\tau
\int\limits_0^{\infty}\D\tau'\int\limits_0^{\infty}\D\tau''
e^{-\frac{i}{\hbar}eV(\tau+2\tau'+\tau'')}\kappa(\tau,\tau',\tau'')
\label{ac3}
\end{equation}
with
\begin{equation}
\kappa(\tau,\tau',\tau'')=
\langle[[[
\hat{H}_+^{(I)}(t),\hat{H}_+^{(I)}(t-\tau)],
\hat{H}_-^{(I)}(t-\tau-\tau')],\hat{H}_-^{(I)}(t-\tau-\tau'-\tau'')]
\rangle_o\;.
\label{kap1}
\end{equation}
The hat indicates that the operators only carry the time dependence with
respect to the electrodes and the environment. The resolution of the
interlaced commutators and the separation of the environment lead to
\begin{eqnarray}
\kappa&=&-\langle\tilde{H}_-(t-\tau-\tau'-\tau'')
\tilde{H}_-(t-\tau-\tau')\tilde{H}_+(t-\tau)\tilde{H}_+(t)
\rangle_o\nonumber\\
&&\times\langle e^{-i{\mit\Phi}(t-\tau-\tau'-\tau'')}
e^{-i{\mit\Phi}(t-\tau-\tau')}
e^{i{\mit\Phi}(t-\tau)}e^{i{\mit\Phi}(t)}\rangle_o+\mbox{7 further terms}\;.
\label{kap2}
\end{eqnarray}
Now the operators $\tilde{H}_{\pm}$ only possess the time dependence with
respect to the electrodes
\[\tilde{H}_{\pm}(t)=e^{\frac{i}{\hbar}(H_L+H_R)t}H_{\pm}
e^{-\frac{i}{\hbar}(H_L+H_R)t}\;.\]
To simplify matters at first we restrict ourselves
to a low-resistance environment ($R_E\ll R_Q$). This means that the
phase correlation functions can be replaced by $1$. The high-resistance case
is dealt with in Sec.~\ref{sec6}. The calculation of the correlation function
$\kappa$ is rather lengthy and cannot be shown in detail. The operators
$H_{\pm}$ have to be expressed by the elementary operators $c$,
$c^{\dagger}$, $\gamma$ and $\gamma^{\dagger}$ which show an exponential
time dependence:
\begin{eqnarray}
\label{cop}
c^{(\dagger)}_{l,\sigma}(t)&=&e^{\pm\frac{i}{\hbar}\ve_l\cdot t}
c^{(\dagger)}_{l,\sigma}\;,\\
\label{gamop}
\gamma^{(\dagger)}_{r,\sigma}(t)&=&e^{\pm\frac{i}{\hbar}E_r\cdot t}
\gamma^{(\dagger)}_{r,\sigma}\;.
\end{eqnarray}
Using Wick's theorem, the correlations can be expressed by the Fermi
distribution functions $f$ and the different time integrations can be
carried out. By means of Dirac's formula, the real part can be taken and the
relevant contribution coming from the first term in Eq.~(\ref{kap2})
turns out to be
\begin{eqnarray}
\frac{8e\pi}{\hbar}\sum\limits_{ll'\atop rr'}&&
f(\ve_l)f(\ve_{l'})f(-E_r)f(-E_{r'})
u_rv_ru_{r'}v_{r'}2T^{\ast}_{lr}T^{\ast}_{l'-r}T_{lr'}T_{l'-r'}
\delta(-\ve_l-\ve_{l'}-2eV)\nonumber\\
&\times &\left[\frac{1}{E_r-\ve_l-eV}\frac{1}{E_{r'}-\ve_{l'}-eV}+
\frac{1}{E_r-\ve_l-eV}\frac{1}{E_{r'}-\ve_l-eV}\right]\;.
\label{kap3}
\end{eqnarray}
The summation over the spin indices has already been performed. The
derivation of Eq.\ (\ref{kap3}) uses the properties
\[u_{r,\sigma}=u_{r,-\sigma};\qquad v_{r,\sigma}=-v_{r,-\sigma}\;.\]
During the calculation one has to treat lots of terms but only a few of them
are nonvanishing and contribute to the Andreev current. In similar way as
above this analysis can be applied to the other seven terms in Eq.\
(\ref{kap2}).

\section{Andreev current}
\label{sec5}
Further calculations require an assumption of the momentum dependence
of the tunneling matrix elements. With respect to quasiparticle tunneling
the squared absolute values of tunnel matrix elements $|T_{lr}|^2$ emerge
which are approximated by their momentum averages at the Fermi
edges of the electrodes $\langle |T_{lr}|^2\rangle $ usually.
Furthermore, the phenomenological tunnel conductance is defined by
\begin{equation}
G\equiv\frac{2\pi e^2}{\hbar}\nu(0)^22\langle |T_{lr}|^2\rangle\;,
\label{cond}
\end{equation}
where $\nu(0)$ is the density of states at the Fermi level in the normal
conducting state. In contrast to this the product of tunnel matrix elements
in Eq.\ (\ref{kap3}) is much more complicated because it depends crucially
on the nature of the electron motion. Some effort has been done to treat
this in terms of a diffusive transport \cite{ave7,hek1}. However, in
case of sufficiently small junctions we can assume that the picture of
ballistic electron motion is correct which means that the scattering of
electrons can be neglected. We follow Ref.~\cite{hek3} and approximate
the momentum averaged product of tunneling matrix elements
$2\langle T^{\ast}_{lr}T^{\ast}_{l'-r}T_{lr'}T_{l'-r'}\rangle$ in terms of
$2\langle|T_{lr}|^2\rangle\cdot 2\langle|T_{l'r'}|^2\rangle/N$, where $N$ is
a correction factor which depends on the geometry of the junction. It
corresponds to the number of independent current channels.
The additional factors $2$ are due to the spin.
Now, the discrete sums are transformed into integrations where the energy
densities of states of the superconductor and the normal conductor arise.
Finally, the Andreev current is given by the following expression:
\begin{eqnarray}
\langle I\rangle^{(A)}(V)&=&\frac{\hbar G^2{\mit\Delta}^2}{\pi e^3 N}
\int\limits_{-\infty}^{\infty}\D\ve_l\int\limits_{-\infty}^{\infty}\D\ve_{l'}
\int\limits_{-\infty}^{\infty}\D E_r\int\limits_{-\infty}^{\infty}\D E_{r'}
f(\ve_l)f(\ve_{l'})f(-E_r)f(-E_{r'})
\frac{{\mit\Theta}(|E_r|-{\mit\Delta})}{\sqrt{E_r^2-{\mit\Delta}^2}}
\frac{{\mit\Theta}(|E_{r'}|-{\mit\Delta})}{\sqrt{E_{r'}^2-{\mit\Delta}^2}}
\nonumber\\
&&\times\left[\frac{1}{E_r-\ve_l-eV}\frac{1}{E_{r'}-\ve_{l'}-eV}+
\frac{1}{E_r-\ve_l-eV}\frac{1}{E_{r'}-\ve_l-eV}\right]
\delta(-\ve_l-\ve_{l'}-2eV)\;.
\label{ac4}
\end{eqnarray}
At zero temperature the Fermi distribution becomes a theta function
($f(x)={\mit\Theta}(-x)$) and the energy integrations can be carried out.
The resulting formula of the Andreev current
\begin{equation}
\langle I\rangle^{(A)}(V)=\frac{\pi}{2}\frac{\hbar G^2{\mit\Delta}}{e^3 N}
\log\left(\frac{1+\frac{eV}{{\mit\Delta}}}{1-\frac{eV}{{\mit\Delta}}}\right)
\quad\mbox{for}
\quad -\frac{{\mit\Delta}}{e}<V<\frac{{\mit\Delta}}{e}
\label{ac5}
\end{equation}
shows logarithmic singularities at $V=\pm{\mit\Delta}/e$. This is not
surprising because ${\mit\Delta}/e$ corresponds just to the voltage threshold
of quasiparticle tunneling and a singularity of such kind is known from
inelastic co-tunneling\cite{ave5}. The singularity which is an artefact of
perturbation theory shows that still higher order terms are necessary to
describe the crossover between quasiparticle and Andreev tunneling. The
divergence disappears leaving a finite enhancement if a finite lifetime
broadening is taken into account.
This can be done either by hand \cite{kre4} or by performing a
nonperturbative resummation technique \cite{laf2,sch8}. Of course, the
singularity will also be smoothed both due to finite temperatures and
finite environment resistances. A Taylor expansion in terms of the voltage
($V\ll{\mit\Delta}/e$) yields the linear behavior
\begin{equation}
\langle I\rangle^{(A)}(V)\approx\pi\frac{\hbar G^2}{e^2 N}V
\label{ac6}
\end{equation}
and the Andreev conductance
\begin{equation}
G^{(A)}=\frac{\D\langle I\rangle^{(A)}(V)}{\D V}\Bigg|_{V=0}=\quad
\frac{\pi\hbar G^2}{e^2N}=\frac{R_QG^2}{2N}\;.
\label{acond1}
\end{equation}
Note, that $G^{(A)}$ is proportional to $R^{-2}$. Fig.\ \ref{fig1} shows the
Andreev current (\ref{ac5}) and its linear approximation (\ref{ac6}) in
the subgap region of quasiparticle tunneling. The scale of quasiparticle
tunneling beyond the threshold is about several orders of magnitude
(factor $2N/(R_QG)\gg 1$) larger than that of the Andreev tunneling.

\section{High-impedance case}
\label{sec6}
The investigation of the high-impedance case is based on Eq.\ (\ref{kap2}).
Now we have to calculate the phase correlation functions which were
so far replaced by $1$. This can be done in Gaussian approximation, for
instance by generalizing the method presented in Ref.\ \cite{ing2}.
The phase correlation functions occuring in Eq.\ (\ref{kap2}) depend on
the function $J$ which contains the information about the structure
of the environment. See e.~g. Refs.\ \cite{dev1,gra1,ing2}.
In the high-impedance case and at $T=0$ there is a linear behavior
\begin{equation}
J(\tau)=-i\frac{E_c}{\hbar}\tau
\label{jot}
\end{equation}
which leads to the following expression for the phase correlation
function of first term of Eq.\ (\ref{kap2})
\begin{equation}
\langle e^{-i{\mit\Phi}(t-\tau-\tau'-\tau'')}
e^{-i{\mit\Phi}(t-\tau-\tau')}
e^{i{\mit\Phi}(t-\tau)}e^{i{\mit\Phi}(t)}\rangle_o=
e^{i\frac{E_c}{\hbar}(\tau+4\tau'+\tau'')}\;.
\label{phase}
\end{equation}
Now the first term of Eq.\ (\ref{kap2}) reads
\begin{eqnarray}
\frac{8e\pi}{\hbar}\sum\limits_{ll'\atop rr'}&&
f(\ve_l)f(\ve_{l'})f(-E_r)f(-E_{r'})
u_rv_ru_{r'}v_{r'}2T^{\ast}_{lr}T^{\ast}_{l'-r}T_{lr'}T_{l'-r'}
\delta(-\ve_l-\ve_{l'}-2eV+4E_c)\nonumber\\
&\times &\left[\frac{1}{E_r-\ve_l-eV+E_c}\frac{1}{E_{r'}-\ve_{l'}-eV+E_c}+
\frac{1}{E_r-\ve_l-eV+E_c}\frac{1}{E_{r'}-\ve_l-eV+E_c}\right]\;.
\label{kap4}
\end{eqnarray}
Finally, taking all relevant contributions into account, one gets
the Andreev current
\begin{equation}
\langle I\rangle^{(A)}(V)=\frac{\hbar G^2{\mit\Delta}}{e^3 N}\frac{1}{2\pi}
\int\limits_{-\hat{v}}^{\hat{v}}\D z
\left[\int\limits_1^{\infty}\D y
\frac{1}{\sqrt{y^2-1}}\frac{2(y-w)}{(y-w)^2-z^2}\right]^2
\Theta(eV-2E_c)-[V\to -V]\;,
\label{ac7}
\end{equation}
where the abbreviations $w=E_c/{\mit\Delta}$ and
$\hat{v}=eV/{\mit\Delta}-2w$ are used (see Fig.\ \ref{fig2}, solid line).
Eq.\ (\ref{ac7}) is valid in the
interval $-({\mit\Delta}+E_c)/e<V<({\mit\Delta}+E_c)/e$. Of course, the
Coulomb blockade of Andreev tunneling ($V_{bl}^{(A)}=2E_c/e$) should be
smaller than the threshold of quasiparticle tunneling
($({\mit\Delta}+E_c)/e$). This is guaranteed for $E_c<\Delta$.
Though the Andreev current cannot be calculated analytically we are able
to derive the differential conductance $g(V)$ (see Fig.\ \ref{fig3}, solid
line)
\begin{equation}
g(V)=\frac{\D\langle I\rangle^{(A)}(V)}{\D V}=\frac{R_Q G^2}{2N}
\frac{1}{\pi^2}
\left[\frac{\frac{\pi}{2}+\arcsin(w-\hat{v})}{\sqrt{1-(w-\hat{v})^2}}
+\frac{\frac{\pi}{2}+\arcsin(w+\hat{v})}{\sqrt{1-(w+\hat{v})^2}}\right]^2
\Theta(eV-2E_c)+[V\to -V]\;.
\label{diffcond1}
\end{equation}
Hence, the Andreev conductance which is defined as the differential
conductance at the blockade voltage $V_{bl}^{(A)}$ turns out to be
\begin{equation}
G^{(A)}=g(V_{bl}^{(A)})=
\frac{R_Q G^2}{2N} \frac{ \left(1+\frac{2}{\pi}\arcsin w\right)^2}{1-w^2}
=\frac{R_QG^2}{2N} \frac{\left(\frac{4}{\pi}
\arctan \frac{\sqrt{1+w}}{\sqrt{1-w}}\right)^2}{1-w^2}\;.
\label{acond2}
\end{equation}
The result coincides with that of Ref.\ \cite{hek3}.
Eq.\ (\ref{acond1}) is reproduced in the limit $E_c\to 0$.
Fig.\ \ref{fig4} (solid line) shows the Andreev conductance as a function of
$w=E_c/{\mit\Delta}$. In case that the Coulomb blockade $V_{bl}^{(A)}=2E_c/e$
tends to the voltage threshold of quasiparticle tunneling
$({\mit\Delta}+E_c)/e$ the Andreev conductance becomes singular ($w\to 1$).

\section{Approximation of Andreev current}
\label{sec7}
Since the Andreev current (cf.\ Eq.\ (\ref{ac7})) is given by an integral
which cannot be carried out analytically, we want to derive an approximated
expression.
This can be achieved by taking into account only the relevant singularities
of the integrand, namely $1/(1-(w-z)^2)$ and $1/(1-(w+z)^2)$. According to
that, the expresion in squared brackets in Eq.\ (\ref{ac7}) is approximated
in the following way:
\[ [\ldots]^2\approx\frac{a_1}{1-(w-z)^2}+\frac{a_2}{1-(w+z)^2}\;,\]
where $a_1$ and $a_2$ are constants. Now the integration can be carried out
and we get
\begin{equation}
\frac{\hbar G^2{\mit\Delta}}{e^3 N}
\frac{a^{\ast}}{4\pi}
\log\left(
\frac{1+w+\hat{v}}{1+w-\hat{v}}\cdot\frac{1-w+\hat{v}}{1-w-\hat{v}}
\right)
\Theta(\hat{v})-[V\to -V]\;,
\label{ac8}
\end{equation}
where a new constant $a^{\ast}=a_1+a_2$ is introduced. Furthermore,
the factor $(1+w+\hat{v})/(1+w-\hat{v})$ in the argument of the
logarithm can be neglected because it is an irrelevant contribution.
The unknown constant $a^{\ast}$ is determined by the demand
\begin{equation}
G^{(A)}_{\mbox{\scriptsize approx}}\Bigg|_{w=0}=\quad
\frac{\hbar G^2}{e^2 N}\frac{a^{\ast}}{2\pi}\quad\equiv\quad
G^{(A)}\Bigg|_{w=0}=\quad\frac{R_Q G^2}{2N}\;,
\label{acond3}
\end{equation}
which gives $a^{\ast}=2\pi^2$. Hence, the approximated Andreev current
reads
\begin{equation}
\langle I\rangle^{(A)}_{\mbox{\scriptsize approx}}(V)\approx
\frac{R_Q G^2}{2N}\frac{{\mit\Delta}}{2e}
\log\frac{{\mit\Delta}-3E_c+eV}{{\mit\Delta}+E_c-eV}
\Theta(eV-2E_c)-[V\to -V]\;.
\label{ac9}
\end{equation}
We have plotted both the exact and the approximated Andreev current as
functions of the voltage for $2E_c<eV<{\mit\Delta}+E_c$ in Fig.\ \ref{fig2}.
As an example $E_c$ is chosen to be  $E_c={\mit\Delta}/2$.
The approximated differential conductance is given by differentiation
\begin{equation}
g(V)=\frac{R_Q G^2}{2N}
\frac{1-w}{(1-w+\hat{v})(1-w-\hat{v})}\Theta(eV-2E_c)+[V\to -V]
\label{diffcond2}
\end{equation}
and can be compared with the exact expression (\ref{diffcond1}) (see Fig.\
\ref{fig3}).
The approximated Andreev conductance reads
\begin{equation}
G^{(A)}_{\mbox{\scriptsize approx}}=\frac{R_Q G^2}{2N}\frac{1}{1-w}\;.
\label{acond4}
\end{equation}
Fig.\ \ref{fig4} shows both the exact and the approximated
expressions of the Andreev conductance as a function of $w=E_c/{\mit\Delta}$.

\acknowledgments
We would like to thank H.-O. Mller for helpful discussions. This work
was supported by the Deutsche Forschungsgemeinschaft.

\section{Conclusion}
\label{sec8}

In conclusion we have shown how the Andreev current through a
NIS junction can be calculated by means of a nonlinear response approach
basing on the elementary Hamiltonian of quasiparticle
tunneling. This method generates in a straightforward way all kinds of
current contributions including quasiparticle and Andreev tunneling.
For low- and high-resistance environments we have calculated the
voltage dependence of the Andreev current and the differential conductance
explicitly. Finally, a simple approximation in case of
the high-impedance environment has been given.


\bibliographystyle{prsty}


\begin{figure}
\caption{The Andreev current in the low-impedance case
(Eq.\ (\protect\ref{ac5}), solid line) and its linear expansion
(Eq.\ (\protect\ref{ac6}), dashed line) are plotted versus
the voltage in the interval $0<eV/{\mit\Delta}$.
Current and voltage are counted in units of
$R_QG^2{\mit\Delta}/(2Ne)$ and ${\mit\Delta}/e$, respectively.}
\label{fig1}
\end{figure}

\begin{figure}
\caption{The Andreev current in the high-impedance case
(Eq.\ (\protect\ref{ac7}), solid line) and its approximation
(Eq.\ (\protect\ref{ac9}), dashed line)
are plotted versus the voltage in the interval $2E_c<eV<{\mit\Delta}+E_c$
for $E_c={\mit\Delta}/2$. Current and voltage are counted in units of
$R_QG^2{\mit\Delta}/(2Ne)$ and ${\mit\Delta}/e$, respectively.}
\label{fig2}
\end{figure}

\begin{figure}
\caption{The differential conductance (Eq.\ (\protect\ref{diffcond1}),
solid line) and its approximation (Eq.\ (\protect\ref{diffcond2}), dashed
line) are drawn as functions of the voltage for
$2E_c<eV<{\mit\Delta}+E_c$ and $E_c={\mit\Delta}/2$.
Conductance and voltage are counted in units of
$R_Q G^2/(2N)$ and ${\mit\Delta}/e$, respectively.}
\label{fig3}
\end{figure}

\begin{figure}
\caption{The Andreev conductance (Eq.\ (\protect\ref{acond2}), solid line)
and its approximation (Eq.\ (\protect\ref{acond4}), dashed line) are plotted
as functions of $w=E_c/{\mit\Delta}$ for $0<w<1$. Conductance is counted
in units of $R_QG^2/(2N)$.}
\label{fig4}
\end{figure}

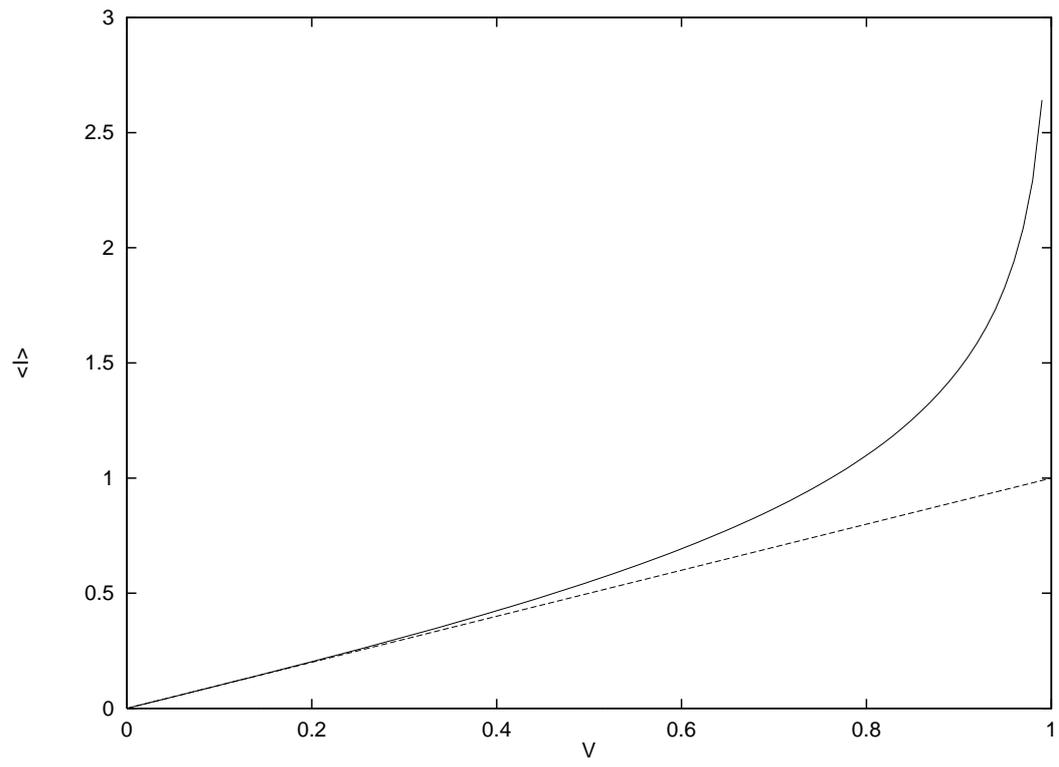
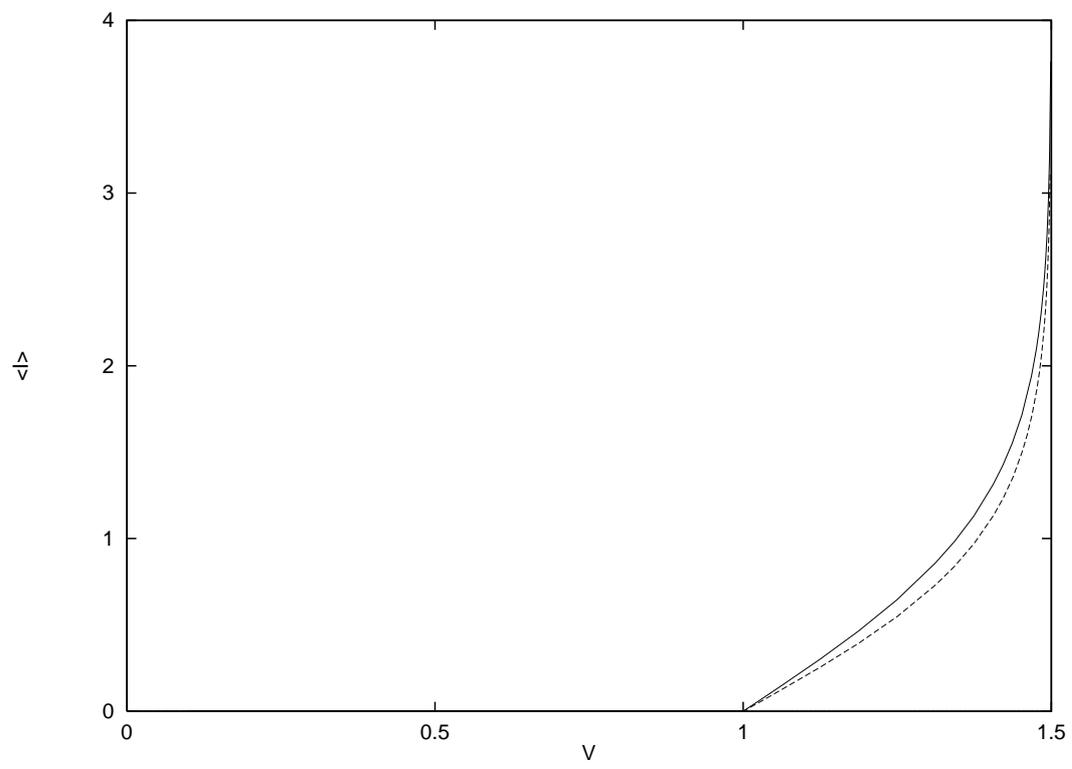
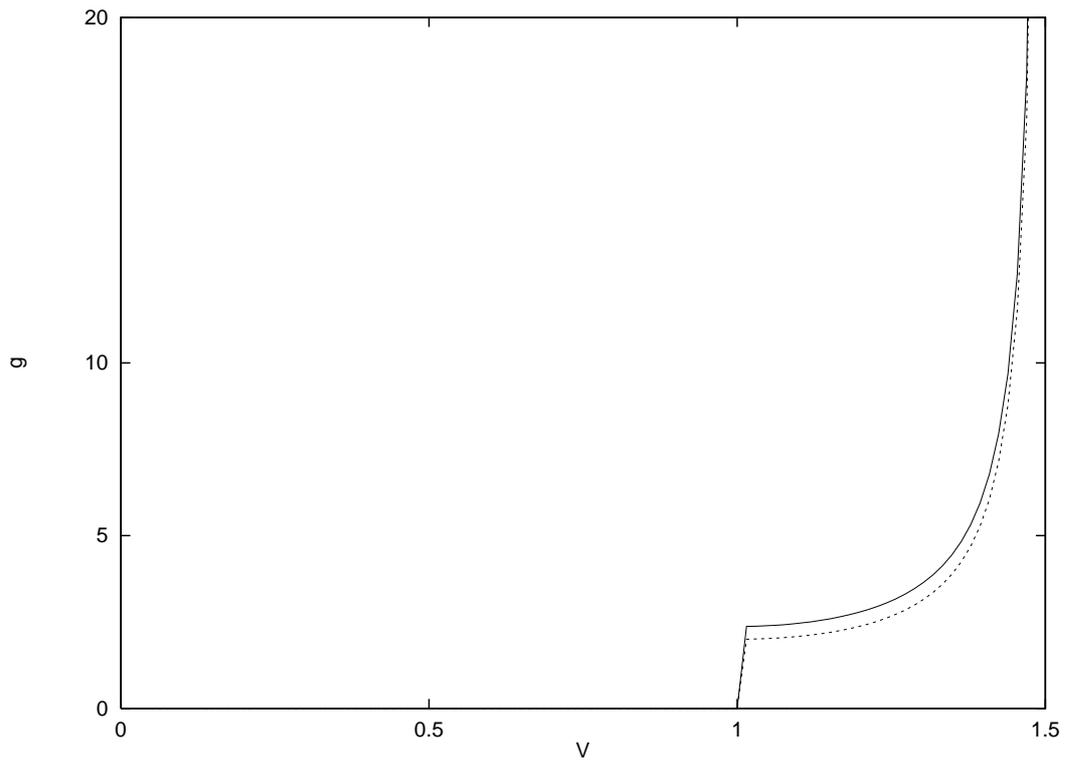
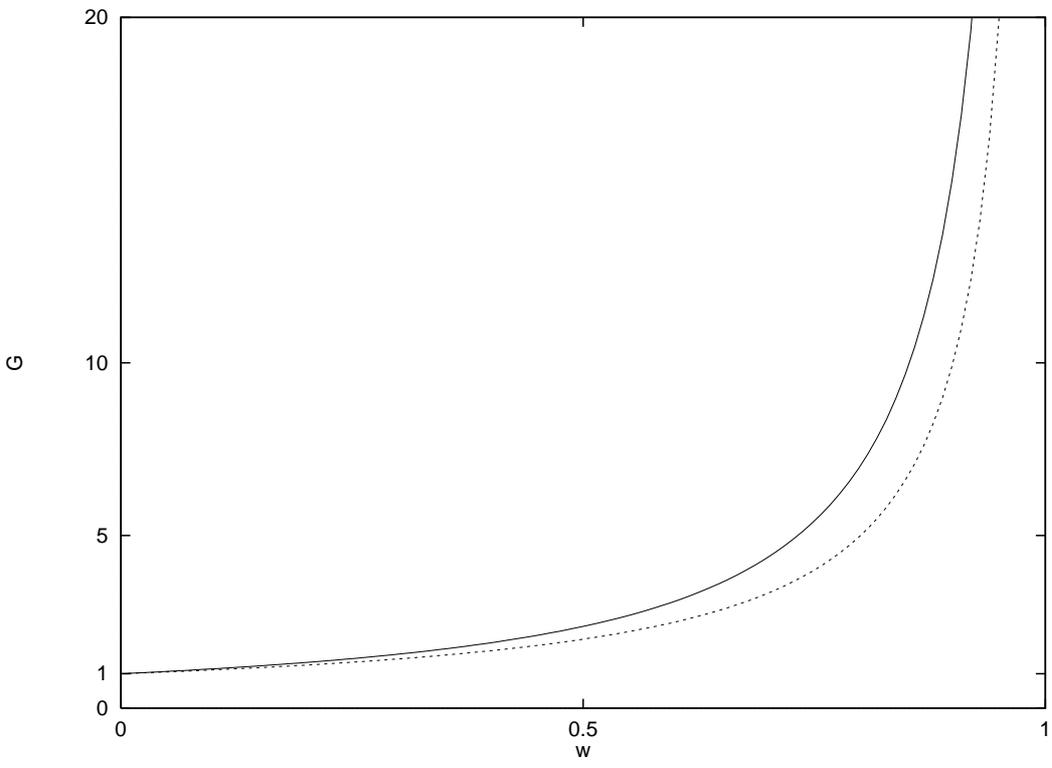
\end{document}